\documentclass[british,english,a4paper,twoside,twocolumn,twocolumn,english,showpacs,preprintnumbers,nofootinbib]{revtex4-1}
\usepackage[T1]{fontenc}
\usepackage[latin9]{inputenc}
\usepackage{fancyhdr}
\usepackage{verbatim}
\usepackage{textcomp}
\usepackage{amsthm}
\usepackage{amsmath}
\usepackage{graphicx}
\usepackage{esint}

\makeatletter
\@ifundefined{textcolor}{}
{%
 \definecolor{BLACK}{gray}{0}
 \definecolor{WHITE}{gray}{1}
 \definecolor{RED}{rgb}{1,0,0}
 \definecolor{GREEN}{rgb}{0,1,0}
 \definecolor{BLUE}{rgb}{0,0,1}
 \definecolor{CYAN}{cmyk}{1,0,0,0}
 \definecolor{MAGENTA}{cmyk}{0,1,0,0}
 \definecolor{YELLOW}{cmyk}{0,0,1,0}
 }

\@addtoreset{equation}{section} 
\renewcommand{\theequation}{\thesection.\arabic{equation}}
\let\old@appendix=\appendix 
\def\appendix{%
   \old@appendix 
   \@addtoreset{equation}{section} 
   \def\theequation{\thesection\arabic{equation}}} %
\fancypagestyle{plain}{\fancyhead{}}
\usepackage{epstopdf}

\makeatother

\usepackage{babel}
\begin{document}

\title{Local conditions separating expansion from collapse in spherically
symmetric models with anisotropic pressures}

\author{José P. Mimoso }

\email{jpmimoso@fc.ul.pt}

\selectlanguage{english}%

\affiliation{Departamento de F\'{\i}sica and Centro de Astronomia e Astrof\'{\i}sica
da Universidade de Lisboa,\\
Faculdade de Ciências, Ed. C8, Campo Grande, 1769-016 Lisboa, Portugal}

\author{Morgan Le Delliou}

\email{Morgan.LeDelliou@uam.es, delliou@cii.fc.ul.pt}

\selectlanguage{english}%

\affiliation{Departamento de Física Matemática, Instituto de Física, Universidade
de São Paulo,\\
CP 66.318 \textemdash{} 05314-970, São Paulo, SP, Brasil}

\altaffiliation{Centro de Astronomia e Astrofísica da Universidade de Lisboa,
Faculdade de Ciências, Ed. C8, Campo Grande, 1769-016 Lisboa, Portugal}

\selectlanguage{english}%

\author{Filipe C. Mena}

\email{fmena@math.uminho.pt}

\selectlanguage{english}%

\affiliation{Centro de Matemática\\
 Universidade do Minho\\
 Campus de Gualtar, 4710-057 Braga, Portugal}


\preprint{
}

\preprint{Version \today}

\date{Received...; Accepted...}
\begin{abstract}
We investigate spherically symmetric spacetimes with an anisotropic
fluid and discuss the existence and stability of a dividing shell
separating expanding and collapsing regions. We resort to a $3+1$
splitting and obtain gauge invariant conditions relating intrinsic
spacetimes quantities to properties of the matter source. We find
that the dividing shell is defined by a generalization of the Tolman-Oppenheimer-Volkoff
equilibrium condition. The latter establishes a balance between the
pressure gradients, both isotropic and anisotropic, and the strength
of the fields induced by the Misner-Sharp mass inside the separating
shell and by the pressure fluxes. This defines a local equilibrium
condition, but conveys also a non-local character given the definition
of the Misner-Sharp mass. By the same token, it is also a generalized
thermodynamical equation of state as usually interpreted for the perfect
fluid case, which now has the novel feature of involving both the
isotropic and the anisotropic stress. We have cast the governing equations
in terms of local, gauge invariant quantities which are revealing
of the role played by the anisotropic pressures and inhomogeneous
electric part of the Weyl tensor. We analyse a particular solution
with dust and radiation that provides an illustration of our conditions.
In addition, our gauge invariant formalism not only encompasses the
cracking process from Herrera and coworkers but also reveals transparently
the interplay and importance of the shear and of the anisotropic stresses.%

\end{abstract}
\maketitle

\section{Introduction}

The universe close to us is inhomogeneous exhibiting structures at
different scales that are the result of the non-linear collapse of
overdensities, and below certain scales these structures seem to be
immune to the overall expansion of the universe. On the other hand,
this picture reveals two different dynamical behaviours that we wish
to %
describe by a global general relativistic solution. This solution
must exhibit expansion on the large scales, and infall at smaller
scales, eventually producing bound structures. It is the understanding
of the interplay between collapsing and expanding regions within the
theory of general relativity (GR) that we aim to address here. This
issue is connected to the general problem of assessing the influence
of global physics into local physics \cite{Ellis:2001cq,Faraoni:2007es},
as well as to the approach to non-perturbative backreaction through
model building \cite{Ellis:2011hk,Clarkson:2011zq}. Another related
problem is that of recollapsing \cite{Barrow1986MNRAS.223..835B,Bondi1969MNRAS.142..333B,Bonnor1985MNRAS.217..597B,Cattoen:2005dx,Burnett1993PhRvD..48.5688B}

In a previous paper \cite{MLeDM09}, we have obtained local conditions
for perfect fluid solutions to collapse within an otherwise cosmologically
expanding background (also in \cite{Delliou:2009dm,Delliou:2009dc}).
We have characterised the locally defined separating shells between
the collapsing and the expanding regions. 

In the present paper we wish to deepen our understanding of the problem
under consideration by overcoming the limits placed by the consideration
of a perfect fluid. While such a description of the matter content
is justified when one deals with an equilibrium configuration, the
consideration of non-equilibrium states requires a more general viewpoint
where anisotropic stresses are present \cite{Mimoso:2011zz}. Indeed,
only for a static equilibrium does one expect to find identical pressures
along all the spatial directions, with the exception being spatially
homogeneous models (in the latter case one might even question whether
this should hold only for the isotropic case). So one should envisage
different directional behaviours, which is precisely what should be
expected both from collapsing or expanding regions within spherically
symmetric models, since the radial and transverse directions behave
differently.

In the present work we investigate spherically symmetric spacetimes
with an anisotropic fluid, but no heat fluxes since we want to concentrate
on the role of the stresses, shear and intrinsic curvature regarding
the problem under consideration. %
We leave the role of heat fluxes for a subsequent work. As in our
previous works \cite{MLeDM09,LeDMM09a} we resort to a $3+1$ splitting,
which allows for a full metric describing both collapse and expanding
regions, and thus avoids having to deal with the matching problem.\textbf{
}We thus make use of a single coordinate patch. This non-perturbative
approach relies on the use of the formalism which has been developed
in a remarkable series of papers by Lasky and Lun using Generalised
Painlevé-Gullstrand (hereafter GPG) coordinates \cite{Adler et al 05,LaskyLun06b,LaskyLun06+}.
We assess the existence and stability of a dividing shell separating
expanding and collapsing regions, in a gauge invariant way. The local
conditions that we find generalize our previous results, and relate
intrinsic spacetimes quantities to quantities characterizing the matter
source. This happens through a generalization of the Tolman-Oppenheimer-Volkoff
equilibrium condition, which, itself, is a generalization of the corresponding
isotropic generalized TOV condition found in \cite{MLeDM09}. Our
condition establishes a relation between the pressure gradients, both
isotropic and anisotropic, and the strength of the fields induced
by the Misner-Sharp mass inside the separating shell and by the pressure
fluxes. This defines a local equilibrium condition, but conveys also
a non-local character given the definitions of the Misner-Sharp mass,
and of the energy function $E$ (see definition in Eq. (\ref{eq:dsLaskyLun})
below). By the same token, it is also a generalized thermodynamical
equation of state as usually interpreted for the perfect fluid case,
which now has the novel feature of involving both the isotropic and
the anisotropic stress.

In addition, this approach has allowed us to express the Einstein
field equations as a dynamical system involving scalar invariants
and local quantities. This formulation reveals the fundamental roles
of combinations of expansion with shear and two sets combining the
electric Weyl with anisotropic stress scalars that are discussed in
their flow evolution, relation to curvature and impact on shear evolution.

To illustrate our results we analyse a particular solution with dust
and radiation. Such a solution stems from the work of Sussman and
Pav\'on \cite{Sussman:1999vx} where, albeit the generality of their
initial formalism, they analysed only the thermodynamic aspects of
the spatially flat spherical solution. We find the conditions caracterising
the matter and radiation content to fulfill the existence of a separating
shell. In turn we also obtain the non-flat elliptic solutions.


On a different context, Herrera and co-workers \cite{Herrera} have
studied small anisotropic perturbations around spherically symmetric
homogeneous fluids in equilibrium. They concluded that this may lead
to instabilities that result in the {}``cracking'' of boundary surface
of compact objects in astrophysics. We recover their results within
our gauge invariant formalism, which not only confirms the important
role of the shear and of the anisotropic stresses but also reveals
transparently their interplay and how they trigger the cracking process.

An outline of the paper is the following: in Section (\ref{sec:splitting-and-gauge})
the GPG formalism of Lasky and Lun and the $3+1$ splitting is revised.
We also define gauge invariant kinematical quantities. In Section
(\ref{sub:General-conditions-defining}) we discuss the existence
of a shell separating collapse from expansion and give general dynamical
conditions. In Section (\ref{sec:Illustration-of-our}) we present
illustrations with a dust plus radiation solution and with the relation
between the separating shell and cracking. Section (\ref{sec:Summary-and-discussion})
gives a discussion of our results.

We shall use $\kappa^{2}=8\pi G$, $c=1$ and the following index
convention: Greek indices $\alpha,\beta,...=1,2,3$ while Latin indices
$a,b,...=0,1,2,3$. 


\section{\label{sec:splitting-and-gauge}$3+1$ splitting and gauge invariants
kinematical quantities}

We set the basic equations in generalised Painlevé-Gullstrand coordinates
following the formalism developed by Lasky and Lun (LL) \cite{LaskyLun06b,LaskyLun06+},
while adapting their derivations for our standpoint which is concerned
with the collapse within an underlying overall expansion, rather than
by collapse on its own.


\subsection{Metric and ADM splitting}

We assume that the flow of the fluid is characterized by the timelike,
normalised 
 vector $n_{a}:=-\alpha\nabla_{a}t=\left[-\alpha,0,0,0\right]$ ($n_{a}n^{a}=-1$),
defining with its lapse $N=\alpha$ and its radial shift vector $N^{\mu}=\left(\beta,0,0\right)$,
and an evolution of the spatially curved three-metric $^{3}g_{\mu\nu}=\text{diag}\left(\frac{1}{1+E},r^{2},r^{2}\sin^{2}\theta\right)$.
Consequently we write the spherically symmetric line element as 
\begin{multline}
ds^{2}=-\alpha\left(t,R\right)^{2}dt^{2}+\frac{1}{1+E\left(t,R\right)}\left(\beta\left(t,R\right)dt+dR\right)^{2}\\
+r\left(t,R\right)^{2}d\Omega^{2},\label{eq:dsLaskyLun}
\end{multline}
which adopts the GPG coordinates of Ref.~\cite{LaskyLun06+} ($d\Omega^{2}:=d\theta^{2}+\sin^{2}\theta d\phi^{2}$).
Notice that the areal radius $r$ differs, in principle, from the
$R$ coordinate to account for additional degrees of freedom that
are required to cope with both a general fluid that includes anisotropic
stresses and heat fluxes. However, in our case we shall ignore the
heat fluxes.%
\footnote{Thus, as discussed below, it becomes possible to restrict to $R=r$.
However, we will\textbf{ }restrain here from doing this identification
to maintain the full generality\textbf{ } in the following equations.%
}

Performing  a ADM 3+1 splitting \cite{EllisElst98,LaskyLun06b,LaskyLun06+}
we use the projection operators along and orthogonal to the flow 
\begin{alignat}{2}
N_{\, b}^{a}:=-n^{a}n_{b} & ,~~~ & h^{ab}:= & g^{ab}+n^{a}n^{b}.\label{eq:ProjNh}
\end{alignat}
 where $h^{ab}$ is the 3-metric on the surface $S_{3}$ normal to
the flow. Those projectors are also used for covariant derivatives:
Along the flow, the proper time derivative of any tensor $X_{\,\, cd}^{ab}$
is 
\begin{align}
\dot{X}_{\,\, cd}^{ab} & :=n^{e}X_{\,\, cd;e}^{ab},
\end{align}
 and in the orthogonal 3-surface, each component is projected with
$h$ (the overbar denotes the covariant derivative and tensor full
orthogonal projection)
\begin{align}
X_{\,\, cd;\bar{e}}^{ab} & :=h_{f}^{a}h_{g}^{b}h_{c}^{i}h_{d}^{j}h_{e}^{k}X_{\,\, ij;k}^{fg}.
\end{align}
 Then the covariant derivative of the flow, from its projections,
is defined as
\begin{multline}
n_{a;b}=N_{b}^{\, c}n_{a;c}+n_{a;\bar{b}}=-n_{b}\dot{n}_{a}+\frac{1}{3}\Theta h_{ab}+\sigma_{ab}\\
+\omega_{ab},\;
\end{multline}
 where the trace of the projection is the expansion of the flow, $\Theta=n_{\,;\bar{a}}^{a}=n_{\,;a}^{a}$,
the rate of shear $\sigma_{ab}$ is its symmetric trace-free part
and its skew-symmetric part is the vorticity $\omega_{ab}$.

On the other hand, we consider an energy-momentum tensor 
\begin{equation}
T^{ab}=\rho\, n^{a}n^{b}+P\, h^{ab}+\Pi^{ab}\;,\label{anisot_stressEMT}
\end{equation}
 where $\rho$ is the energy density, $P$ is the pressure and $\Pi^{ab}$
is the anisotropic stress tensor. $\Pi^{ab}n_{b}=0$ and ${\Pi^{a}}_{a}=0$,
i.e., the anisotropic stress $\Pi^{ab}$ is orthogonal to $n^{a}$
and traceless.

The spherical symmetry implies that all the quantities $X_{\alpha\beta}={h_{\alpha}}^{a}{h_{\beta}}^{a}\, X_{ab}$
share the same spatial eigen-directions characterised by the traceless
3-tensor ${P^{\alpha}}_{\beta}={\rm {diag}\left[-2,1,1\right]}$,
such that 
\begin{equation}
X_{\alpha\beta}=w(t,R)\, P_{\alpha\beta}\;.
\end{equation}
In fact, one can define in 4 dimensions the traceless projector as
$P_{ab}=h_{ab}-h_{\, c}^{c}\frac{\dot{n}_{a}\dot{n}_{b}}{\dot{n}_{d}\dot{n}^{d}}$,
admitting from the bulk all the properties of the 3-projector in the
hypersurfaces. Because of spherical symmetry, we then can decompose
any spatial 2-tensor into its trace and traceless parts: $X_{ab}=h_{a}^{\, c}h_{b}^{\, d}X_{cd}=X\frac{h_{ab}}{3}+\xi P_{ab}$,
with its trace being $X=X_{\, a}^{a}$ and its traceless eigenvalue
being $\xi$. Therefore, we have the following decompositions for
traceless quantities:
\begin{itemize}
\item For the anisotropic stress 
\begin{equation}
\Pi_{ab}=\Pi(t,R)\, P_{ab}
\end{equation}

\item For the shear tensor (traceless extrinsic curvature) 
\begin{equation}
\sigma_{ab}=a(t,R)\, P_{ab}
\end{equation}

\item For the trace-free, 3-dimensional Riemann tensor, which measures the
departures from constant spatial curvature 
\begin{equation}
^{\left(3\right)}R_{\alpha\beta}-\frac{1}{3}\,^{\left(3\right)}g_{\alpha\beta}\,^{\left(3\right)}R=q(t,R)\, P_{\alpha\beta}
\end{equation}

\item For the trace-free Hessian of the lapse function 
\begin{equation}
\frac{1}{\alpha}\,\left(D_{\gamma}D_{\mu}-\frac{1}{3}\,^{\left(3\right)}g_{\gamma\mu}D^{\beta}D_{\beta}\right)\alpha=\epsilon(t,R)\, P_{\gamma\mu}
\end{equation}

\item For the electric part of the Weyl tensor we have 
\begin{equation}
E_{ab}=\Sigma(t,R)\, P_{ab}.
\end{equation}

\end{itemize}
There is no magnetic part of the Weyl tensor due to the spherical
symmetry \cite{Stephani:2003tm}, and thus the models fall into the
class of that has been dubbed \emph{silent} universes \cite{Barnes:1989ah,Bruni:1994nf}.
Another consequence of the spherical symmetry is that the flow is
irrotational, $\omega_{ab}=0$.

\subsection{The Einstein Field Equations}


It is well known that the ADM approach separates the ten Einstein's
Field Equations (EFE) into four constraints on the hypersurfaces and
six evolution equations. Spherical symmetry reduces them to 2+2.

The EFE can then be written as a set of propagation equations: %
the trace and tracefree%
\footnote{Note the sign differences in front of the Lie derivatives terms compared
with \cite{LaskyLun06b}; otherwise the Raychaudhuri equation restricted
to the FLRW case does not get the usual sign for $\dot{H}$.%
} orthogonal contractions of the EFE, the double orthogonal contracted
and flow projected the Bianchi identity (once with the tracefree orthogonal
projector)%
\footnote{In other terms, contracted with $h_{\, d}^{b}n^{a}P_{\, e}^{c}$,
the Bianchi identities yield the Weyl evolution.%
} read 
\begin{align}
-2\mathcal{L}_{n}\Theta= & \frac{^{3}R}{2}+\Theta^{2}+9a^{2}-\frac{2}{\alpha}D^{\mu}D_{\mu}\alpha+3\kappa^{2}P-3\Lambda,\label{eq:evol1}\\
\mathcal{L}_{n}a= & -a\Theta+\epsilon-q+\kappa^{2}\,\Pi\label{eq:evol2}\\
\mathcal{L}_{n}\Sigma= & -\frac{\kappa^{2}}{2}\mathcal{L}_{n}\Pi-\frac{\kappa^{2}}{2}\left(\rho+P-2\Pi\right)a\nonumber \\
 & -\left(3\Sigma+\frac{\kappa^{2}}{2}\Pi\right)\left(\frac{\Theta}{3}+a\right).\label{eq:WeylEvol}
\end{align}
They are accompanied with spacelike constraints: the gauge invariant
radial balance, which proceeds from the cross projection of the EFE,
and the tidal forces, obtained from  the double orthogonal contracted,
acceleration projected Bianchi identity (again, once with the tracefree
orthogonal projector)%
\footnote{Or more precisely, the Bianchi identities contracted with $h_{\, e}^{c}\dot{n}^{b}P_{\, d}^{a}$,
so they yield the Weyl constraint.%
} yield 
\begin{align}
\left(\frac{\Theta}{3}+a\right)^{\prime} & =-3a\frac{r^{\prime}}{r},\label{eq:TrjnhEFE:MomentumConstr}\\
\frac{4\pi}{3}\left(\rho+3\Pi\right)^{\prime} & =-\Sigma^{\prime}-3\left(\Sigma+\frac{\kappa^{2}}{2}\Pi\right)\frac{r^{\prime}}{r}.\label{eq:WeylConstraint}
\end{align}
Finally, the Hamiltonian constraint reads, in the presence of a cosmological
constant,
\begin{align}
^{\left(3\right)}R+\frac{2}{3}\Theta^{2}-6a^{2}= & 2\kappa^{2}\rho+2\Lambda.\label{eq:Hamiltonian}
\end{align}
From the twice contracted Bianchi identities we also derive, along
and orthogonal to the flow,
\begin{align}
\mathcal{L}_{n}\rho= & -\Theta(\rho+P)-6\Pi\, a\label{eq:nBianchi}\\
0= & \left(D^{k}+\dot{n}^{k}\right)\left(\Pi_{ik}+h_{ik}P\right)+\left[\rho-\left(P-2\Pi\right)\right]\dot{n}_{i}\nonumber \\
 & -n_{i}\left[\Theta P+6\Pi a\right].\label{eq:hBianchi}
\end{align}
The latter equation gives the heat fluxes evolution \cite{LaskyLun06+},
which we set to zero here, since we are restricting our analysis to
the case where these fluxes are absent. We thus have 
\begin{equation}
0=-\left(\rho+P-2\Pi\right)\frac{\alpha^{\prime}}{\alpha}-\left(P-2\Pi\right)^{\prime}+6\Pi\frac{r^{\prime}}{r}.\label{eq:jCons}
\end{equation}
 The inspection of the system of equations (\ref{eq:evol1}-\ref{eq:jCons})
thus tells us that the anisotropic stress shows up in all but Eqs.
(\ref{eq:evol1}), (\ref{eq:TrjnhEFE:MomentumConstr}), and (\ref{eq:Hamiltonian}).
This reveals the importance of the anisotropic pressures in explicitly
contributing to the evolution of the shear, the electric part of the
Weyl tensor, and of the lapse function $\alpha$ \cite{Chan:1992zz,Mimoso:1993ym,Coley:1994yt,McManus:1994ys,Herrera,Herrera:2004xc,Herrera:2008bt}. 

It is worth noticing at this point that we have included the cosmological
constant $\Lambda$ for the sake of completeness. However, none of
the results that follow will depend on its presence. Indeed, we can
without loss of generality make $\Lambda=0$, or alternatively absorb
it into $\rho$ and $P$.%

Introducing the Misner-Sharp mass \cite{MisnerSharp} and following
\cite{LaskyLun06+} 
\begin{equation}
M^{\prime}=\frac{\kappa^{2}}{2}\rho r^{2}r^{\prime}\label{MS_mass}
\end{equation}
 it is possible to derive%
\footnote{By analogy with the perfect fluid case, it is also possible to derive
\begin{align}
r^{\prime}\mathcal{L}_{n}E & =2\left(1+E\right)\left[-\left(\mathcal{L}_{n}r\right)^{\prime}-\frac{\beta^{\prime}}{\alpha}r^{\prime}\right],\label{eq:LieE}
\end{align}
\begin{gather}
\mathcal{L}_{n}M=-\frac{\kappa^{2}}{2}r^{2}\left(P-2\Pi\right)\mathcal{L}_{n}r.\label{eq:LieM}
\end{gather}
} 
\begin{equation}
\left({\mathcal{L}_{n}r}\right)^{2}=\frac{2M}{r}+(1+E)\,\left(r^{\prime}\right)^{2}-1+\frac{1}{3}\Lambda r^{2}\label{eq:LieR}
\end{equation}
 and 
\begin{equation}
-{\mathcal{L}_{n}^{2}\, r}=\frac{M}{r^{2}}+\frac{\kappa^{2}}{2}(P-2\Pi)r-\left(1+E\right)\frac{\alpha^{\prime}}{\alpha}r^{\prime}-\frac{1}{3}\Lambda r\;.\label{eq:Lie2r}
\end{equation}
 This allows us to extend the generalization of the TOV function made
in \cite{MLeDM09} to the case where anisotropic stresses are present:
\begin{equation}
\mathrm{gTOV}=-{\mathcal{L}_{n}^{2}\, r}\;.\label{eq:gTOVdef}
\end{equation}
 Since, in the absence of heat fluxes we have 
\begin{equation}
-\frac{\alpha^{\prime}}{\alpha}=\frac{1}{(\rho+P-2\Pi)}\,\left[\left(P-2\Pi\right)^{\prime}-6\Pi\frac{r^{\prime}}{r}\right],
\end{equation}
then Eqs. (\ref{eq:Lie2r}) and (\ref{eq:gTOVdef}) become
\begin{multline}
\mathrm{gTOV}=-{\mathcal{L}_{n}^{2}\, r}=\frac{M}{r^{2}}+\frac{\kappa^{2}}{2}(P-2\Pi)r\\
+\frac{\left(1+E\right)r^{\prime}}{(\rho+P-2\Pi)}\,\left[\left(P-2\Pi\right)^{\prime}-6\Pi\frac{r^{\prime}}{r}\right]-\frac{1}{3}\Lambda r.\label{eq:gTOV}
\end{multline}
This tells us that, when going from the isotropic perfect fluid to
the case of an anisotropic content in the above equations, we have
to replace $P$ by $P-2\Pi$ and introduce an extra term related to
anisotropic stresses.

\section{General conditions defining a shell separating expansion from collapse\label{sub:General-conditions-defining}}

We now derive the generalized local conditions for the existence of
a separating shell at $r=r_{\star}$. First, we require a stationarity
condition on the shell 
\begin{equation}
\left({\mathcal{L}_{n}r}_{\star}\right)^{2}=\frac{2M_{\star}}{r_{\star}}+(1+E_{\star})\,\left(r_{\star}^{\prime}\right)^{2}-1+\frac{\Lambda}{3}r_{\star}^{2}=0\;,\label{cond1_TP}
\end{equation}
 and second, we need an equilibrium condition to be satisfied on the
shell 
\begin{equation}
-{\mathcal{L}_{n}^{2}\, r_{\star}}=\frac{M_{\star}}{r_{\star}^{2}}+\frac{\kappa^{2}}{2}(P_{\star}-2\Pi_{\star})r_{\star}-\left(1+E_{\star}\right)\frac{\alpha_{\star}^{\prime}}{\alpha_{\star}}r_{\star}^{\prime}-\frac{\Lambda}{3}r_{\star}=0\;.\label{gTOV2}
\end{equation}
Indeed, from Eq. (\ref{eq:gTOV}), the $\mathrm{gTOV}_{\star}=0$
equation of state for the stationarity of the separating shell becomes
now 
\begin{multline}
-\frac{1}{(\rho+P-2\Pi)_{\star}}\,\left[\left(P-2\Pi\right)^{\prime}-6\Pi\frac{r^{\prime}}{r}\right]_{\star}\\
=\left[\frac{\frac{M}{r^{2}}+\left[\frac{\kappa^{2}}{2}(P-2\Pi)-\frac{1}{3}\Lambda\right]r}{1-2\frac{M}{r}-\frac{1}{3}\Lambda r^{2}}\right]_{\star}r_{\star}^{\prime}\;.\label{anis_gTOV}
\end{multline}

Thus the existence of a spherical shell separating an expanding outer
region from an inner region collapsing to the center of symmetry,
depends essentially on two conditions.%
\footnote{We emphasize again that $\Lambda$ is written here for the sake of
generality but is not required for the conditions to hold.%
} The former (\ref{cond1_TP}) amounts to the vanishing of the kinetic
energy of the shell, and establishes the precise balance between the
analogues of the total and potential energies at the dividing shell.
The latter condition (\ref{gTOV2}), combined with the former (\ref{cond1_TP}),
is the generalization of the TOV equation for the present case, and
is necessary for the equilibrium of the shell. There are noticeable
differences with respect to the original problem in the form of the
TOV equation \cite{Tolman:1939jz,Oppenheimer:1939ne}. The isotropic
pressure gradient $P'$ is replaced by $\left(P-2\Pi\right)^{\prime}$,
the gravitational mass $\rho+P$ is consistently traded into $(\rho+P-2\Pi)$,
and there is a new additional term,$-6\Pi\frac{r^{\prime}}{r}$, involving
the anisotropic stress $\Pi$ and hence reflecting its additional
contribution to the balance of pressures and forces per unit mass.
It is worth stressing that our result does not rely on the assumption
of a static equilibrium of the spherical distribution of matter, and
consequently does not assume that all the internal spherical shells
are constrained to satisfy the TOV equation. Here the generalized
TOV equation is just satisfied at the dividing shell. On the neighboring
shells it won't be satisfied, and these shells will either be collapsing
or expanding since they are not in equilibrium.%
\footnote{We won't consider here the possible case where the inner shells move
outwards and the outer shells move inwards, so that shell crossing
results. Here we are just interested in characterizing the converse
situation where the inner and outer shells depart. The occurrence
of shell crossing in inhomogeneous models with anisotropic pressures
is discussed in \cite{AndrzejKrasinski:1997zz}.%
} Moreover, the generalized TOV function depends on the spatial 3-curvature
in a more general way than the original TOV function.

It goes without saying that, from the conditions (\ref{cond1_TP})
and (\ref{gTOV2}), it is straightforward to realize that the absence
of pressure gradients between the neighboring shells prevents the
existence of a separating shell in the spatially homogeneous FLRW
models. %

Since we have 
\begin{align}
\left(\frac{\Theta}{3}+a\right) & =\frac{{\mathcal{L}_{n}r}}{r}\label{H_r}\\
\mathcal{L}_{n}\left(\frac{\Theta}{3}+a\right)+\left(\frac{\Theta}{3}+a\right)^{2} & =\frac{{\mathcal{L}_{n}^{2}r}}{r}
\end{align}
 we see that the turning point condition (\ref{cond1_TP}) does not
imply necessarily the vanishing of the expansion nor of the shear,
but it rather means that these quantities should satisfy $\Theta_{\star}=-3a_{\star}$
at the separating shell $r=r_{\star}$ as there
\begin{align}
\Theta_{\star}+3a_{\star}= & 0\label{eq:LimitDef}\\
\mathcal{L}_{n}\left(\frac{\Theta}{3}+a\right)_{\star}= & 0.\label{eq:gTOV0}
\end{align}
If one of $\Theta$ or $a$ were to vanish at this locus we would
then have the other quantity vanishing as well. This limit case corresponds
to the total staticity of the separating shell.%

\subsection{The  relation in dynamics between non-local and local conditions}

Although the conditions (\ref{cond1_TP}) and (\ref{anis_gTOV}) that
characterize the separating shell hold locally, at $r=r_{\star}$,
they involve non-local quantities, namely $M$ and $E$. Indeed, from
the construction of $M$, Eq. (\ref{MS_mass}), we see that the profile
of the distribution of matter inside the separating shell is taken
into account.

It is however possible to find local conditions involving local, rather
than non-local quantities and this is addressed in what follows.

Given Eq. (\ref{H_r}) it is possible to relate the condition (\ref{eq:LieR})
to the Hamiltonian constraint (\ref{eq:Hamiltonian}) that generalizes
the Friedman equation. With that purpose, we recast the latter (also
known as the Gau\ss -Codazzi equation, obtained from $\frac{1}{3}n^{a}n^{b}G_{ab}$)
as 
\begin{equation}
\left(\frac{\Theta}{3}+a\right)^{2}=\frac{\kappa^{2}}{3}\rho-\frac{^{\left(3\right)}R}{6}+\frac{\Lambda}{3}+2a\,\left(\frac{\Theta}{3}+a\right)\;,\label{eq:Gauss-Codazzi}
\end{equation}
 so that we conclude that 
\begin{equation}
\frac{2M}{r^{3}}+(1+E)\,\left(\frac{r^{\prime}}{r}\right)^{2}-\frac{1}{r^{2}}=\frac{\kappa^{2}}{3}\rho-\frac{^{\left(3\right)}R}{6}+2a\,\left(\frac{\Theta}{3}+a\right)\;.
\end{equation}
 In parallel, we also wish to clarify the relation between the gTOV
function, expressed with the gauge invariant of Eq. (\ref{H_r}),
\begin{align}
\mathrm{gTOV}= & -r\,\left(\mathcal{L}_{n}\left(\frac{\Theta}{3}+a\right)+\left(\frac{\Theta}{3}+a\right)^{2}\right)
\end{align}
 and the \textquotedbl{}generalized\textquotedbl{} Raychaudhuri equation,
obtained from contracting the Ricci identity with the combination
of projectors $-\frac{1}{6}\left(2h^{ac}+P^{ac}\right)n^{b}$,
\begin{align}
{\mathcal{L}_{n}\left(\frac{\Theta}{3}+a\right)}+\left(\frac{\Theta}{3}+a\right)^{2}= & \epsilon+\frac{1}{3\alpha}D^{k}D_{k}\alpha-\frac{\kappa^{2}}{6}\,(\rho+3P)\nonumber \\
 & -\left(\Sigma-\frac{\kappa^{2}}{2}\Pi\right)+\frac{\Lambda}{3}\;.\label{eq:gRay}
\end{align}
It is interesting to relate $^{\left(3\right)}R$ to $E$ from its
metric expression (\ref{eq:3Rdef}) 
\begin{gather}
\frac{^{\left(3\right)}R}{2}=-(1+E)\,\left(\frac{r^{\prime}}{r}\right)^{2}+\frac{1}{r^{2}}-2\frac{\sqrt{1+E}}{r}\left(\sqrt{1+E}r^{\prime}\right)^{\prime}.
\end{gather}

We see that the separating conditions (\ref{cond1_TP}) and (\ref{anis_gTOV})
now translate into (from Eq. \ref{eq:Gauss-Codazzi}) 
\begin{equation}
\frac{^{\left(3\right)}R}{2}|_{r_{\star}}=\kappa^{2}\rho|_{r_{\star}}+\Lambda\;,\label{c1:gFried=00003D0}
\end{equation}
 and (from Eq. \ref{eq:gRay}) 
\begin{align}
-\epsilon|_{r_{\star}}-\frac{1}{3\alpha}D^{k}D_{k}\alpha|_{r_{\star}}= & -\frac{\kappa^{2}}{6}\,(\rho+3P)|_{r_{\star}}\nonumber \\
 & -\left(\Sigma-\frac{\kappa^{2}}{2}\Pi\right)|_{r_{\star}}+\frac{\Lambda}{3}\;.\label{c2:gRay=00003D0}
\end{align}

The former of these equations reveals that the stationarity condition
requires $^{\left(3\right)}R>0$, when $\rho,\,\Lambda>0$.%
\footnote{Strictly speaking, when $\kappa^{2}\rho_{\star}+\Lambda>0$.%
} It no longer explicitly involves the Misner-Sharp mass $M_{\star}$,
but just the local energy density $\rho_{\star}$ . The latter condition
emerges from the generalized Raychaudhuri equation (\ref{eq:gRay})
and, besides involving local quantities defined at $r=r_{\star}$
as well, it reveals that the important role of the pressure gradient
of Eq. (\ref{anis_gTOV}) is now translated by the Hessian trace and
traceless eigenvalue on the left-hand side of Eq. (\ref{c2:gRay=00003D0}).

\subsection{Non-locality around the shell}

It is possible to express the expansion scalar $\Theta$ in terms
of the areal radius and its Lie and radial derivatives: 
\begin{equation}
\Theta=\left(\frac{\left(\mathcal{L}_{n}r\right)^{\prime}}{r^{\prime}}+2\frac{\mathcal{L}_{n}r}{r}\right)\;,\label{theta_of_Lnr}
\end{equation}
 and from it to derive 
\begin{equation}
\left(r^{2}\mathcal{L}_{n}r\right)^{\prime}=\Theta r^{2}r^{\prime}\;.\label{theta_of_Lnr_b}
\end{equation}
This expression reveals that, in the inhomogeneous spherical models,
the expansion scalar $\Theta$ is not just the logarithmic derivative
of the spatial volume along the timelike flow, unlike what happens
in the spatially homogeneous Friedman-Lemaître-Robertson-Walker (FLRW)
models. Indeed, we see that it rather contains the logarithmic Lie
derivative along the flow of the areal radius $r$ and of its radial
gradient $r^{\prime}$.

From (\ref{theta_of_Lnr_b}), upon integration and %
choosing a fixed fiducial areal radius $r_{0}$ defined as $r_{0}=r(t,R_{0}(t))=cst$,
we obtain 
\begin{equation}
\mathcal{L}_{n}r=\frac{1}{r^{2}}\,\int_{r_{0}}^{r}\,\Theta\, r^{2}\,{\rm d}r+\frac{1}{r^{2}}\,\left[r^{2}\mathcal{L}_{n}r\right]_{r_{0}}.\label{eq:NonLocalTheta}
\end{equation}
This result shows that the turning point condition at $r_{\star}$
yields 
\begin{align}
-\left[r^{2}\mathcal{L}_{n}r\right]_{r_{0}} & =\int_{r_{0}}^{r_{\star}}\,\Theta\, r^{2}\,{\rm d}r.\label{eq:NonLocalThetaStar}
\end{align}
The integral on the right-hand side vanishes if the initial parameter
$\left[r^{2}\mathcal{L}_{n}r\right]_{r_{0}}$ vanishes at some interior
value $r_{0}<r_{\star}$. This requires the vanishing of the expansion
$\Theta$ at some intermediate value of $r$, $r_{0}<\tilde{r}<r_{\star}$,
since it has to change signs within the interval of integration (we
assume that no shell crossing occurs in that range). Differentiating
equation (\ref{eq:NonLocalTheta}) with respect to the flow, we obtain
\begin{align}
\mathcal{L}_{n}^{2}r= & -\frac{2\mathcal{L}_{n}r}{r^{3}}\,\left(\int_{r_{0}}^{r}\,\Theta\, r^{2}\,{\rm d}r+\left[r^{2}\mathcal{L}_{n}r\right]_{r_{0}}\right)\nonumber \\
 & +\frac{1}{r^{2}}\,\left\{ \mathcal{L}_{n}\left(\int_{r_{0}}^{r}\,\Theta\, r^{2}\,{\rm d}r\right)+\mathcal{L}_{n}\left[r^{2}\mathcal{L}_{n}r\right]_{r_{0}}\right\} \nonumber \\
= & \left[\mathcal{L}_{n}r\right]\left\{ \Theta-\frac{2}{r}\,\left[\mathcal{L}_{n}r\right]\right\} \nonumber \\
 & +\frac{1}{r^{2}}\,\left\{ \int_{r_{0}}^{r}\,\frac{\partial\Theta}{\partial\tau}\, r^{2}\,{\rm d}r+\mathcal{L}_{n}\left[r^{2}\mathcal{L}_{n}r\right]_{r_{0}}\right\} \nonumber \\
= & -\mathrm{gTOV},\label{eq:NonLocalTheta_gTOV}
\end{align}
where $\tau$ denotes proper time%
. %
This is the equation that generalizes the Eq. (3.27) of \cite{MLeDM09}
and that corresponds to Eq. (21) of di Prisco \emph{et al.} \cite{DiPrisco94}.
It corroborates once again the claim of a non-locality of the radial
acceleration. From Eq.~(\ref{eq:NonLocalTheta}) we realise that
this non-locality is inherent in the radial expansion, and is already
present in the energy condition defining $r_{\star}$ Eqs.~(\ref{cond1_TP},\ref{eq:LimitDef})
and in our $\mathrm{gTOV}$ condition Eqs.~(\ref{eq:gTOVdef},\ref{eq:Lie2r},\ref{gTOV2}),
since both implicate $M$ which is an integral between $0$ and $r_{\star}$. 

From the previous equations (\ref{eq:NonLocalTheta}) and (\ref{eq:NonLocalTheta_gTOV})
we see that at the separating shell we have
\begin{align}
-\mathcal{L}_{n}\left[r^{2}\mathcal{L}_{n}r\right]_{r_{0}} & =\int_{r_{0}}^{r_{\star}}\,\frac{\partial\Theta}{\partial\tau}\, r^{2}\,{\rm d}r\label{eq:NonLocalTheta_gTOVstar}
\end{align}
 which means that the integral on the right-hand side vanishes if
the term $-\mathcal{L}_{n}\left[r^{2}\mathcal{L}_{n}r\right]_{r_{0}}$
vanishes at an interior value $r_{0}<r_{\star}$. This shows that
the vanishing of the proper time derivative of the expansion $\mathcal{L}_{n}\Theta$
occurs then at some intermediate value between $r_{0}$ and $r_{\star}$.
In the case when $\mathcal{L}_{n}\left[r^{2}\mathcal{L}_{n}r\right]_{r_{0}}=0$
at the center, we recover the result of Di Prisco \emph{et al.} \cite{DiPrisco94},
establishing the vanishing of the radial acceleration, i.e.  $\mathcal{L}_{n}\Theta=0$,
at some $0<r<r_{\star}$.\textbf{ }However this results is derived
here in a non perturbative way, and in a more general way than in
Ref.~\cite{DiPrisco94}.

\subsection{Dynamics around the shell}

In this section, we will address the dynamics of the system under
consideration, adding various restrictions of interest for the rest
of the paper and, in each case, examining the dynamics of the matter-trapped
shell.

\subsubsection{Dynamical system of the imperfect fluid\label{sub:dynamical-system-of}}

\paragraph{Governing equations}

The dynamical system of partial differential equations (PDE's) that
results from the 3+1 splitting and the use of the local kinematical
and geometric quantities is given by Eqs. (\ref{eq:gRay}, \ref{eq:evol2},
\ref{eq:WeylEvol}, \ref{eq:Gauss-Codazzi}, \ref{eq:jCons}, \ref{eq:TrjnhEFE:MomentumConstr},
\ref{eq:WeylConstraint}) and a constraint (Eq.~\ref{eq:EFEricciConst})
on the Weyl tensor induced by the differences in the shear equation
obtained by projections both from the Einstein field equations (Eq.~\ref{eq:evol2}
from $\frac{P^{dc}}{6}G_{cd}$) and from the Ricci identities (Eq.~\ref{eq:shearRicci-1}
from the projection$-\frac{1}{6}P^{ac}n^{b}$). We restate the whole
system as
\begin{align}
\mathcal{L}_{n}\left(\frac{\Theta}{3}+a\right)= & \frac{1}{3\alpha}D^{k}D_{k}\alpha+\epsilon-\left(\frac{\Theta}{3}+a\right)^{2}\nonumber \\
 & -\frac{\kappa^{2}}{6}\left(\rho+3\left(P-2\Pi\right)\right)+\frac{\Lambda}{3}-\left(\Sigma+\frac{\kappa^{2}}{2}\Pi\right),\label{gRayGen}\\
\mathcal{L}_{n}a= & -a\,\Theta+a\,\left(\frac{\Theta}{3}+a\right)\nonumber \\
 & +\left[\epsilon-\left(\Sigma-\frac{\kappa^{2}}{2}\Pi\right)\right],\label{eq:shearRicci-1}
\end{align}
 
\begin{align}
\mathcal{L}_{n}\left(\Sigma+\frac{\kappa^{2}}{2}\Pi\right)= & -\frac{\kappa^{2}}{2}a\left(\rho+P-2\Pi\right)\nonumber \\
 & -\left[2\left(\Sigma+\frac{\kappa^{2}}{2}\Pi\right)+\left(\Sigma-\frac{\kappa^{2}}{2}\Pi\right)\right]\times\nonumber \\
 & \times\left(\frac{\Theta}{3}+a\right),\label{eq:LieofWeyl}\\
\Sigma+\frac{\kappa^{2}}{2}\Pi= & q+a\,\left(\frac{\Theta}{3}+a\right),\label{eq:EFEricciConst}\\
\left(\frac{\Theta}{3}+a\right)^{2}= & \frac{\kappa^{2}}{3}\rho-\frac{^{\left(3\right)}R}{6}+\frac{\Lambda}{3}+2a\,\left(\frac{\Theta}{3}+a\right)\;,\\
\left(P-2\Pi\right)^{\prime}= & 6\Pi\frac{r^{\prime}}{r}-\left(\rho+P-2\Pi\right)\frac{\alpha^{\prime}}{\alpha},\\
\left(\frac{\Theta}{3}+a\right)^{\prime}= & -3a\frac{r^{\prime}}{r},\\
\frac{\kappa^{2}}{6}\rho^{\prime}= & -\frac{\left(\left(\Sigma+\frac{\kappa^{2}}{2}\Pi\right)r^{3}\right)^{\prime}}{r^{3}}\;.\label{eq:WeylConstraintSimple}
\end{align}
In this formulation the equations reveal%
\footnote{Notice that Eq. (\ref{gRayGen}) can also be noted
\begin{align}
\mathcal{L}_{n}\left(\frac{\Theta}{3}+a\right)= & \frac{1}{3\alpha}D^{k}D_{k}\alpha+\epsilon-\left(\frac{\Theta}{3}+a\right)^{2}\nonumber \\
 & -\frac{\kappa^{2}}{6}\left(\rho+3P\right)+\frac{\Lambda}{3}-\left(\Sigma-\frac{\kappa^{2}}{2}\Pi\right).\label{gRayGen-1}
\end{align}
} the fundamental role played by some combinations of gauge invariant
quantities like expansion and shear,  electric Weyl and anisotropic
stress. In the latter case, they emerge in two different combinations
that play different and important roles in the governing equations,
as we will discuss in what follows. $\Sigma+\frac{\kappa^{2}}{2}\Pi$
acts as a source for density inhomogeneities as seen in Eq. (\ref{eq:WeylConstraintSimple}).
From Eq. (\ref{eq:EFEricciConst}) we see that this is related to
the 3-curvature distortion of the hypersurfaces as well as to the
distortion of the extrinsic curvature, as expected. The role of the
other combination is clearly revealed in the shearfree subsection
\ref{sub:Dynamical-system-restricted} that follows.

Alternatively, one can present Eq. (\ref{eq:TrjnhEFE:MomentumConstr})
in a form parallel to that of Eq. (\ref{eq:WeylConstraintSimple})
\begin{align}
\frac{\Theta^{\prime}}{3}+\frac{\left(ar^{3}\right)^{\prime}}{r^{3}}= & 0.\label{eq:constraint_2alt}
\end{align}

\paragraph{Dynamics of the shell}

On the separating shell, the dynamics can be expressed from the EFE
and Bianchi identities. It takes the form of the residual constraint
from the Raychaudhuri equation (Eq. \ref{eq:evol1}+\ref{eq:Hamiltonian}/2)/6
\begin{align}
-\mathcal{L}_{n}\frac{\Theta_{\star}}{3}-\left(\frac{\Theta_{\star}}{3}\right)^{2}+\frac{1}{3\alpha_{\star}}D^{\mu}D_{\mu}\alpha_{\star}= & \frac{\kappa^{2}}{6}\left(\rho_{\star}+3P_{\star}\right)-\frac{\Lambda}{3},
\end{align}
and the \textquotedbl{}generalized\textquotedbl{} Raychaudhuri Eq.
(\ref{eq:gRay})
\begin{align}
\epsilon_{\star}+\frac{1}{3\alpha_{\star}}D^{\mu}D_{\mu}\alpha_{\star}= & \frac{\kappa^{2}}{6}\,\left(\rho_{\star}+3P_{\star}\right)+\left(\Sigma_{\star}-\frac{\kappa^{2}}{2}\Pi_{\star}\right)-\frac{\Lambda}{3}.
\end{align}
The Hamiltonian constraint yield the local curvature of the shell,
$^{3}R_{\star}=2\kappa^{2}\rho_{\star}+2\Lambda$, the momentum constraint
governs the expansion and shear transfer across the shell, $\left(\frac{\Theta}{3}+a\right)_{\star}^{\prime}=-3a_{\star}\frac{r_{\star}^{\prime}}{r_{\star}}$,
the Weyl constraint from the shear equations links it directly to
the 3-curvature residual\textbf{
\begin{align}
\Sigma_{\star}+\frac{\kappa^{2}}{2}\Pi_{\star}= & q_{\star}\;,
\end{align}
}the density remains conserved by Eq. (\ref{eq:nBianchi}) which,
with Eq. (\ref{eq:LimitDef}), reads now
\begin{align}
\mathcal{L}_{n}\rho_{\star}= & -\Theta_{\star}(\rho+P-2\Pi)_{\star}.
\end{align}
The Eq. (\ref{eq:jCons}) gives a part of the gTOV staticity condition\textbf{.
}The Weyl constraint Eq.(\ref{eq:WeylConstraint}) governs the balance
of anisotropic stress and energy density across the shell. But most
interestingly, the evolution of the electric part of the Weyl tensor
is bound to that of the anisotropic stress by  Eq.(\ref{eq:LieofWeyl})
which reduces here to%
\begin{align}
\mathcal{L}_{n}\left(\Sigma+\frac{\kappa^{2}}{2}\Pi\right)_{\star}= & \frac{\kappa^{2}}{6}\left(\rho+P-2\Pi\right)_{\star}\Theta_{\star}.
\end{align}
{} This is to be related with the studies on cracking by Herrera et
al. \cite{DiPrisco94,Herrera,Herrera:2004xc,Herrera:2008bt}. We now
restrict to shear free flows.

\subsubsection{Dynamical system restricted to shear free flows\label{sub:Dynamical-system-restricted}}

We set out to restrict to shear free flows as they constitute an important
subcase in many studies, i.e. as in \cite{Barnes:1989ah,Herrera:2010ed,Ellis:2011hk,Ellis:2011pi,Senovilla:1997bw}
or even in cosmological FLRW models.

\paragraph{Governing equations}

The Eq. (\ref{eq:evol2}) %
reveals the necessary and sufficient condition for a shear-free flow
to be 
\begin{equation}
\epsilon-\left(\Sigma-\frac{\kappa^{2}}{2}\Pi\right)=0\;.\label{eq:ShearRicci-shearFree}
\end{equation}
Here the combination of electric Weyl and anisotropic stress that
govern the shear evolution appears clearly in its role. This generalizes
the result of \cite{Mimoso:1993ym} to the case of non-vanishing acceleration.

The remaining equations of the system (assuming \ref{eq:ShearRicci-shearFree})
that differ from the general case reduce to 
\begin{align}
\mathcal{L}_{n}\Theta= & \frac{1}{\alpha}D^{k}D_{k}\alpha-\frac{\Theta^{2}}{3}-\frac{\kappa^{2}}{2}\left(\rho+3P\right)+\Lambda\;,\label{gRay_3_shearfree}\\
\mathcal{L}_{n}\left(\Sigma+\frac{\kappa^{2}}{2}\Pi\right)= & -\left[2\left(\Sigma+\frac{\kappa^{2}}{2}\Pi\right)+\left(\Sigma-\frac{\kappa^{2}}{2}\Pi\right)\right]\frac{\Theta}{3},\label{eq:LieOfWeyl-shearfree}\\
\Sigma+\frac{\kappa^{2}}{2}\Pi= & q,\label{eq:EFEricciConst-shearfree}\\
\frac{\Theta^{2}}{3}= & \kappa^{2}\rho-\frac{^{3}R}{2}+\Lambda\;,\\
\Theta^{\prime}= & 0\;.\label{eq:constraint_2-shearfree}
\end{align}

Notice that Eq. (\ref{eq:EFEricciConst-shearfree}) shows that in
the shearfree case, the \textbf{$\Sigma+\frac{\kappa^{2}}{2}\Pi$}
combination of\textbf{ }electric Weyl and anisotropic stress relates
only to the 3-curvature distortion of the hypersurfaces. It also is
a generalization of the constraint $\kappa^{2}\Pi=q=2\Sigma$ found
in \cite{Mimoso:1993ym,McManus:1994ys,Coley:1994yt}. Moreover,\textbf{
}Eq. (\ref{eq:LieOfWeyl-shearfree}) that governs its evolution can
be re-expressed, using Eqs. (\ref{eq:EFEricciConst-shearfree}) and
(\ref{eq:ShearRicci-shearFree}), as
\begin{align*}
\mathcal{L}_{n}q= & -\left[2q+\epsilon\right]\frac{\Theta}{3},
\end{align*}
so $q$ is damped by $\frac{2\Theta}{3}$. Therefore the sign of $\Theta$
determines the increase or decrease of the 3-curvature distortion.
Expansion dampens the distortion while collapse enhances it%
.\textbf{ }More importantly Eq. (\ref{eq:constraint_2-shearfree})
implies that $\Theta$ does not depend on $R$, and therefore 
\begin{equation}
\alpha\,\mathcal{L}_{n}\Theta=\frac{\partial\Theta}{\partial t}=D^{k}D_{k}\alpha-\alpha\,\left[\frac{\Theta^{2}}{3}+\frac{\kappa^{2}}{2}\left(\rho+3P\right)-\Lambda\right]=\chi(t)\;,\label{gRay_3_shearfree_b}
\end{equation}
 where $\chi$ is a function of just the time coordinate $t$. The
Hessian trace is thus determined by the Friedman acceleration sources
and a time dependent term: $\frac{1}{\alpha}D^{k}D_{k}\alpha=\frac{\kappa^{2}}{2}\,\left(\rho+3P\right)-\Lambda+\frac{\Theta^{2}(t)}{3}+\frac{\chi(t)}{\alpha}$.

\paragraph{Dynamics of shear free limiting shell}

Since at $r_{\star}$ we further have $\Theta_{\star}=0$, the remaining
changed equations of the system reduce at that locus to 
\begin{align}
\alpha_{\star}\mathcal{L}_{n}\Theta_{\star}=\frac{\partial\Theta_{\star}}{\partial t}= & D^{k}D_{k}\alpha_{\star}-\alpha_{\star}\frac{\kappa^{2}}{2}\left(\rho+3P\right)_{\star}+\alpha_{\star}\Lambda\nonumber \\
 & =-3\alpha_{\star}\mathcal{L}_{n}a_{\star}=0\;,\label{gRay_3_shearfree_star}\\
\mathcal{L}_{n}\left(\Sigma+\frac{\kappa^{2}}{2}\Pi\right)_{\star}= & 0\;,\label{geomet_constr_shearfree_star}\\
\Theta_{\star}^{\prime}= & 0\;.
\end{align}
 From eq. (\ref{geomet_constr_shearfree_star}) we realize then that
$\left(\Sigma+\frac{\kappa^{2}}{2}\Pi\right)_{\star}=q_{\star}$ is
a constant of the motion along the flow $n^{a}$ (timelike vector
fields). 

In the shear-free case the expansion scalar throughout is only a function
of time, and the relation between the local values of the electric
part of the Weyl tensor and of the anisotropic stress does not change
along the orbits of the shells. It is also worth noticing that the
3-curvature of the dividing shell is completely determined by the
local energy density.

A \textquotedbl{}limit\textquotedbl{} case is the case of a static
initial configuration $\Theta=0$  in addition to the vanishing of
the shear. We will consider this case in the subsection on the cracking
phenomena.

\subsubsection{Dynamical system restricted to geodesic flow}

The following case of geodesic flow is defined by no acceleration.
The perfect fluid solutions are dealt with in \cite{Herlt1996GReGr..28..919H}.%
\footnote{There, spherically symmetric, inhomogeneous models are studied, and
it is shown that there are no exact, geodesic, non-static perfect
fluid solutions with nonzero shear.%
} This implies $\alpha'=0$, and therefore $\epsilon=0$, $D^{k}D_{k}\alpha=0$,
and it is advisable to set $\alpha(t)=1$, and use gLTB coordinates
(generalisation from the Lemaître-Tolman-Bondi, hereafter LTB, coordinates).
As a remark, the more restrictive geodesic, shear-free flows are subject
to 
\begin{equation}
\frac{\kappa^{2}}{2}\Pi=\Sigma\label{shearfree_geodesic}
\end{equation}
recovering the Mimoso and Crawford result \cite{Mimoso:1993ym}, and
the subsequent discussion of Coley and McManus \cite{Coley:1994yt,McManus:1994ys}.

\paragraph{Governing equations}

The equations for geodesic flows that differ from the general case
now reduce to
\begin{align}
\mathcal{L}_{n}\left(\frac{\Theta}{3}+a\right)= & -\left(\frac{\Theta}{3}+a\right)^{2}\nonumber \\
 & -\left\{ \frac{\kappa^{2}}{6}\left(\rho+3P\right)+\left(\Sigma-\frac{\kappa^{2}}{2}\Pi\right)\right\} +\frac{\Lambda}{3},\label{gRay_5}\\
\mathcal{L}_{n}a= & -a\,\Theta+a\,\left(\frac{\Theta}{3}+a\right)-\left(\Sigma-\frac{\kappa^{2}}{2}\Pi\right),\label{eq:shearRicci_5}
\end{align}
 
\begin{align}
\left(P-2\Pi\right)^{\prime}= & 6\Pi\frac{r^{\prime}}{r}\;.\label{eq:constraint_1_5}
\end{align}
This case is interesting as it corresponds to the generalisation of
the classic LTB model \cite{Lemaitre33,Tolman:1939jz} as well as
to the Sussman and Pavon \cite{Sussman:1999vx,Sussman:2008kr} example
we will use later. The major difference from the general case is the
absence of Hessian trace, $\frac{1}{\alpha}D^{a}D_{a}\alpha$, and
traceless Hessian, $\epsilon$, in the equations (\ref{gRay_5}),
(\ref{eq:shearRicci_5}) and (\ref{eq:constraint_1_5}). In particular
Eq. (\ref{eq:constraint_1_5}) displays a completely different radial
constraint: not only the inertial mass is no longer involved, as the
acceleration vanishes, but also in this way the anisotropic stress
is the only source for the inhomogeneity of the pressure. For a perfect
fluid the pressure should be spatially homogeneous, as found in \cite{Herlt1996GReGr..28..919H}.

\paragraph{Dynamics of geodesic limiting shells}

Further restricting to the shell $r_{\star}$ we have the remaining
changed equations 
\begin{align}
\mathcal{L}_{n}\left(\frac{\Theta}{3}+a\right)_{\star}=0= & -\left\{ \frac{\kappa^{2}}{6}\left(\rho+3\left(P-2\Pi\right)\right)+(\Sigma+\frac{\kappa^{2}}{2}\Pi)\right\} _{\star}\nonumber \\
 & +\frac{\Lambda}{3},\label{gRay_5_star}\\
\mathcal{L}_{n}a_{\star}= & \frac{\Theta_{\star}^{2}}{3}-\left(\Sigma-\frac{\kappa^{2}}{2}\Pi\right)_{\star},\label{eq:shearRicci_5_star}\\
\left(P-2\Pi\right)_{\star}^{\prime}= & 6\Pi_{\star}\frac{r_{\star}^{\prime}}{r_{\star}},\\
\left(\frac{\Theta}{3}+a\right)_{\star}^{\prime}= & \Theta_{\star}\frac{r_{\star}^{\prime}}{r_{\star}}\;.\label{eq:constraint_2_5_star}
\end{align}

The definition of the matter-trapped shell then implies Eq. (\ref{gRay_5_star})
which is the local version of the gTOV, i.e., the local $\mathrm{gRAY}=0$
equation. If in addition we have the shear free condition (\ref{shearfree_geodesic}),
we see that the matter-trapped shell imposes that $(\rho+3P)$ be
locally constant or vanishing (if $\Lambda=0$)%
. On the other hand Eq. (\ref{eq:constraint_2_5_star}) shows that
the value of $\left(\frac{\Theta}{3}+a\right)$ in the neighborhood
of the separating shell is non vanishing and that $\left(\frac{\Theta}{3}+a\right)'_{\star}>0$
provided $\Theta_{\star}r_{\star}'>0$.

\paragraph{Geodesic Misner-Sharp mass and electric Weyl}

For the geodesic flow, we have, from Eqs. (\ref{gTOV2}) and (\ref{eq:gRay}),
the latter in the form of Eq. (\ref{gRay_5}), a relation between
the Misner-Sharp mass and the electric Weyl: 
\begin{equation}
\frac{M}{r^{3}}=\left\{ \frac{\kappa^{2}}{6}\left[\rho+3\Pi\right]-\Sigma\right\} .
\end{equation}

\subsection{Separation and expansion}

In cosmology, the expansion of the background universe is understood
as the condition on the universal fluid flow of $\Theta>0$. In Sec.
\ref{sub:General-conditions-defining}, we have extended the definition
of \cite{MLeDM09} for matter-trapped surfaces separating expansion
from collapse, however it should be explicited that, because of its
definition (\ref{eq:LimitDef}), the expansion of the outside region
does not precisely cover the usual expansion region: the separating
shell itself can have non-zero expansion and thus one of the said
collapsing or expanding region may contain the $\Theta=0$ shell.
However, the choice was not laid on such shell because the present
definition yields the staticity condition on that surface (Eq. \ref{anis_gTOV})
which is not, in general the case for turnaround shells ($\Theta=0$).

The meaning of expansion in the terms of Sec. \ref{sub:General-conditions-defining}
is linked with the areal radius: the luminosity distance of a shell
to the centre. Thus the static shell keeps its luminosity distance
to the centre while expanding regions appear so in the luminosity
distance space.

Isolating the Ricci curvature of spatial hypersurfaces $^{\left(3\right)}R$
in Eqs. (\ref{eq:Hamiltonian}) and (\ref{eq:Gauss-Codazzi}), however,
reveals that both $\Theta=0$ and Eq. (\ref{eq:LimitDef}) require
$^{\left(3\right)}R>0$, placing the respective surfaces both in the
positively curved region of spacetime, where the region between the
two must lie. In models with negatively curved regions, those expanding
shells will therefore be contained in the expansion regions defined
for both expansion scalar and areal radius. The flat and closed background
can still present expansion infinities in both senses, as seen in
\cite{LeDMM09a}, although the general treatment can be more complex.

\section{\label{sec:Illustration-of-our}Applications of our results }

\subsection{Sussman-Pav\'on exact solution: radiation and matter}

To illustrate our results we turn our attention to an exact solution
derived by Sussman and Pav\'on for a spherically symmetric model
with a matter content consisting of a combination of dust and radiation
that exhibits anisotropic stress, but no heat fluxes and $\Lambda=0$
\cite{Sussman:1999vx}.

In order to do that we need to translate the metric (\ref{eq:dsLaskyLun})
in the LTB form used in \cite{Sussman:1999vx}, following their assumption
of comoving, i.e. geodesic, flow (here the velocity of light $c$
is reintroduced)
\begin{align*}
n_{SP}^{a} & =c\delta_{t}^{a}.
\end{align*}
That judiciously imposed condition translates into a flow without
acceleration, which leads in terms of metric components to $\alpha=\alpha\left(t\right)$.
Thus, the time function can always be rescaled to absorb the lapse
$\alpha\dot{t}=c$.

\subsubsection{Coordinate transform from GPG to LTB}

Canceling the metric crossed term gives a similar relation than the
perfect fluid generalised LTB formulation of \cite{LaskyLun06b},
while the radial term imposes 
\begin{align}
\left(\beta\dot{t}+\dot{R}\right)R^{\prime} & =0,\label{eq:Comdrdt}\\
R^{\prime2} & =r^{\prime2},\label{eq:Comdr2}
\end{align}
(with $R=R_{GPG}$, the $^{\prime}$ and $\dot{}$ denoting derivatives
in the gLTB frame). Compared with \cite{LaskyLun06+}, the absence
of heat flux suggests that the extra degree of freedom provided by
a spacetime dependent lapse in the GPG frame becomes superfluous.
The new areal radius, from Eq. (\ref{eq:Comdr2}) is reset to the
GPG radial coordinate $r=R_{GPG}=\mathcal{R}$ (here for convenience
we change notation for the GPG radial coordinate), yet is still spacetime
dependent. Then, as in the perfect fluid case in \cite{LaskyLun06b},
the coordinate transform (\ref{eq:Comdrdt}) is such that $\beta dt+d\mathcal{R}\propto dR$.
Taking $t(T)=c\int\frac{dT}{\alpha}$ and $r(T,R)$, we have then
the condition 
\begin{equation}
\beta\partial_{T}t+\partial_{T}\mathcal{R}=0,\label{eq:CondLaskyLTB}
\end{equation}
 which becomes (in GPG coordinates)
\begin{align}
\frac{\beta}{\alpha}= & -\frac{\dot{\mathcal{R}}}{c}.\label{eq:betaRdot}
\end{align}
Moreover, Eq. (\ref{eq:LieE}) implies that, in the new gLTB coordinates,
$E=E(R)$. Consequently, the line element (\ref{eq:dsLaskyLun}) can
be rewritten as 
\begin{equation}
ds^{2}=-c^{2}dT^{2}+\frac{\left(\partial_{R}r\right)^{2}}{1+E(R)}dR^{2}+r^{2}d\Omega^{2},\label{eq:dsLTB}
\end{equation}
as in \cite{Sussman:1999vx}.

\subsubsection{Restricted dynamical equations}

The crucial anstaz adopted by Sussman and Pav\'on was the assumption
that the flow is geodesic, keeping as close as possible to the case
where dust is the only component present, i.e., as in the original
LTB case. The Bianchi contracted identity (\ref{eq:hBianchi}), together
with the geodesic condition $\alpha^{\prime}=0\Leftrightarrow\dot{n}^{a}=0$,
imply that 
\begin{align}
D^{b}\left(h_{ib}P+\Pi_{ib}\right)-n_{i}\left[\Theta P+6\Pi a\right] & =0,
\end{align}
 that is, in gLTB coordinates, 
\begin{equation}
(P-2\Pi)^{\prime}-6\Pi\,\frac{r^{\prime}}{r}=0,
\end{equation}
 where the prime stands for differentiation with respect to the geodesic
$R$. For practical purposes this amount to have 
\begin{equation}
M_{dust}=M(R)\;,
\end{equation}
 and 
\begin{equation}
M_{rad}=\frac{W(R)r_{i}(R)}{2r(T,R)}\;,
\end{equation}
 so that Eq.(\ref{eq:LieR}), including $c$, now reads 
\begin{align}
\dot{r}^{2}= & c^{2}\left(2\frac{M}{r}+\frac{Wr_{i}}{r^{2}}+E\right).\label{eq:RadEvolLTB}
\end{align}

From Eq. (\ref{eq:RadEvolLTB}), one is led to the following solution%
\footnote{The elliptic integral, from Eq. (\ref{eq:RadEvolLTB}), reads $\pm c\int\,{\rm d}t=\int\frac{rdr}{\sqrt{Er^{2}+2Mr+Wr_{i}}}=\int\frac{dY}{2\sqrt{E}\sqrt{Y+cst}}-\frac{M}{E^{3/2}}\int\frac{dX}{\sqrt{X^{2}+cst}}$.%
}, generalised from \cite{Sussman:1999vx} to encompass the cases where
$E\neq0$, 
\begin{multline}
\pm c\int\,{\rm d}t=\left\{ \frac{\sqrt{Er^{2}+2Mr+Wr_{i}}}{E}-\right.\\
\left.-\frac{M}{E^{3/2}}\,\ln\left(\frac{E^{1/2}\sqrt{Er^{2}+2Mr+Wr_{i}}+Er+M}{E}\right)\right\} _{r_{i}}^{r}\;.
\end{multline}

\subsubsection{Existence of a separating shell in the generalised Sussman-Pavón
solutions}

We see that the vanishing of the right-hand side of Eq. (\ref{eq:RadEvolLTB})
provides one of the conditions for the dividing shell, while the other
condition that corresponds to the gTOV equation will be derived from
the radial acceleration 
\begin{equation}
\frac{\ddot{r}}{c^{2}}=-\frac{M}{r^{2}}-\frac{Wr_{i}}{r^{3}},\label{eq:RadAccLTB}
\end{equation}
or directly from combining Eqs. (\ref{eq:evol1}/6+\ref{eq:evol2})
in the gLTB frame with Eq. (\ref{eq:RadEvolLTB}). So we find that
\begin{align}
\frac{Wr_{i}}{2r^{3}} & =\frac{\kappa^{2}}{2}\left(P-2\Pi\right)r.\label{eq:RayLTB}
\end{align}
The form of the gTOV condition in the gLTB frame (Eqs. \ref{eq:gTOVdef},
\ref{eq:Lie2r} with conditions $\alpha^{\prime}=0$) can then be
recognised in Eq. (\ref{eq:RadAccLTB}) using the Raychaudhuri constraint
(\ref{eq:RayLTB}).

The existence of a matter-trapped shell in the solution inspired by
\cite{Sussman:1999vx} requires both Eqs. (\ref{eq:RadEvolLTB}) and
(\ref{eq:RadAccLTB}) to be zero on some $r=r_{\star}$. This implies
the existence of $r{}_{\star}$ and, from Eq. (\ref{eq:RadAccLTB}),
and then Eq. (\ref{eq:RayLTB}), 
\begin{align}
W_{\star}= & -M_{\star}\frac{r_{\star}}{r_{i\star}} & \Rightarrow W(R_{\star}) & <0 & \Leftrightarrow P_{\star} & <2\Pi_{\star}.\label{eq:MTcondition}
\end{align}
This latter condition shows that, in order to allow the separating
shell to exist locally, the model of Sussman and Pavón\textbf{ }\cite{Sussman:1999vx}\textbf{
}must contain regions where the transverse pressures balance the radial
pressure. This can be understood with Eq. (\ref{eq:RayLTB})$\times r^{2}$:
the radiation Misner-Sharp mass corresponds to the flux of the pressures
across the shell. Setting Eq. (\ref{eq:RadEvolLTB}) to zero implies,
using again Eq. (\ref{eq:RayLTB}), 
\begin{align}
E_{\star}= & -2\frac{M_{\star}}{r_{\star}}-\frac{W_{\star}r_{i\star}}{r_{\star}^{2}}=-2\frac{M_{\star}}{r_{\star}}-\kappa^{2}\left(P_{\star}-2\Pi_{\star}\right)r_{\star}^{2},
\end{align}
while Eq. (\ref{eq:RadAccLTB}) gives
\begin{align}
r_{\star}= & \sqrt[3]{\frac{M_{\star}}{\kappa^{2}\left(2\Pi_{\star}-P_{\star}\right)}}
\end{align}
so with (\ref{eq:RayLTB}), the energy/curvature parameter $E$ reads
\begin{align}
E_{\star}= & -\frac{M_{\star}}{r_{\star}}=-\sqrt[3]{\kappa^{2}\left(2\Pi_{\star}-P_{\star}\right)}M_{\star}^{\frac{2}{3}}=\frac{M_{\star}^{\frac{2}{3}}W_{\star}^{\frac{1}{3}}r_{i\star}^{\frac{1}{3}}}{r_{\star}^{\frac{4}{3}}}<0.
\end{align}
Again, as in the perfect fluid case \cite{MLeDM09}, the separating
shell only exists in elliptic regions ($E<0$). Finally with Eq. (\ref{eq:MTcondition})
we have
\begin{align}
r_{\star}= & -\frac{W}{M}r_{i},
\end{align}
and thus
\begin{align}
E_{\star}= & \frac{M^{2}}{Wr_{i}}.
\end{align}
For outward initial flows, this requires $r_{i}\le r_{\star}$, thus
the additional condition $W\le-M<0$.

\subsubsection{Dynamical analysis and global shell}

\begin{figure}
\includegraphics[width=1\columnwidth]{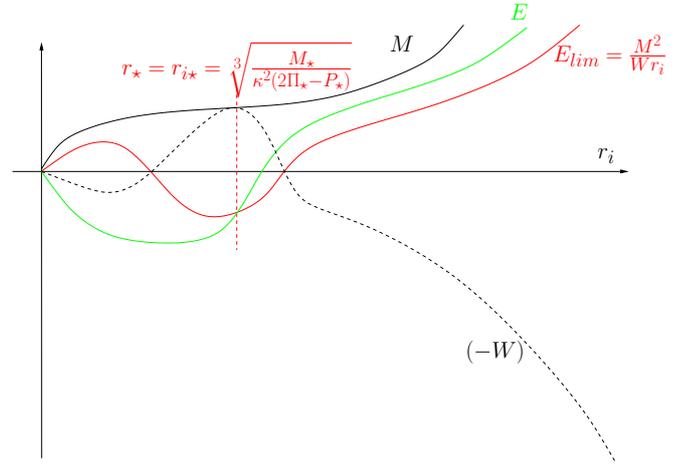}

\caption{\label{fig:Global-analysis-yielding}Global, qualitative, analysis
yielding the separating shell at the intersection of $E$ with $E_{lim}$.
Following \cite{LeDMM09a}, we construct initial cosmological conditions
with a $W<0$ region such that Eqs.~(\ref{eq:RadEvolLTB}) and (\ref{eq:RadAccLTB})
are 0 simultaneously, that is $E=E_{lim}$ and $W=-M$, so the intersection
of the curves gives a global dynamical separation. }

\end{figure}

As for the examples of Ref. \cite{MLeDM09}, a dynamical analysis
\begin{figure}
\includegraphics[width=1\columnwidth]{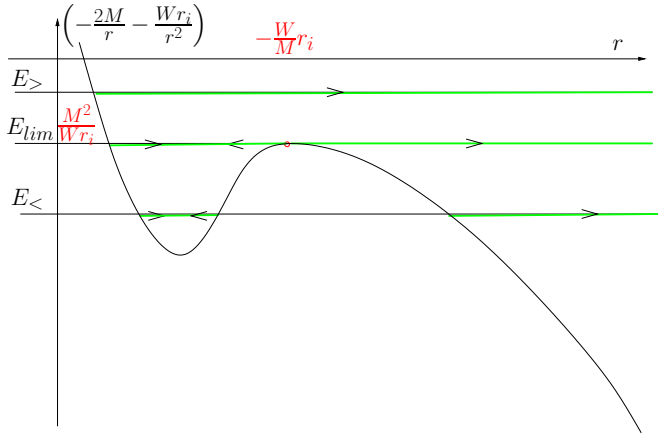}

\caption{\label{fig:Dynamical-analysis-of}Dynamical analysis of a local $W<0$
shell. The dynamic for a given shell (fixed $M$, $W$ and $E$ without
shell crossing) obeys Eq. (\ref{eq:RadEvolLTB}). It then behaves
as a one dimensional particle in an effective potential, following
\cite{LeDMM09a}. We draw a qualitative energy diagram to illustrate
the definition of the critical curvature/energy $E_{lim}$, when it
lies in a region of $W<0$. The various cases of $E_{>}>E_{lim}$,
$E_{<}<E_{lim}$and $E=E_{lim}$ yield unbound, bound and marginally
bound behaviours.}

\end{figure}
 of Eq. (\ref{eq:RadEvolLTB}) can be performed in the regions where
$W<0$ (see Fig. \ref{fig:Dynamical-analysis-of}), required by Eq.
(\ref{eq:MTcondition}). Initial conditions with cosmological outwards
initial areal radius velocity flow, FLRW outer behaviour ($M,W\underset{r\rightarrow\infty}{\sim}r^{3}$,
$E\underset{r\rightarrow\infty}{\sim}r^{2}$) and an intermediate
$W\le-M<0$ region can be qualitatively obtained (Fig.~\ref{fig:Global-analysis-yielding}).
Then using $E_{lim}=\frac{M^{2}}{Wr_{i}}$, and choosing initial velocities
such as $E$ crosses $E_{lim}$ in the $W<0$ region, one gets the
global separation (see Fig. \ref{fig:Global-analysis-yielding}).
Although we allow for a region where the Misner-Sharp mass of the
radiation fluid is negative, we remind that only the total density
of the fluid is actually meaningful and point out that we should keep
\begin{align*}
M^{\prime}+W^{\prime}= & \frac{\kappa^{2}}{2}r_{i}^{2}r_{i}^{\prime}\left(\rho_{m}+\rho_{r}\right)_{i}\ge0, & \textrm{so }M+W\ge & 0.
\end{align*}
This implies that only a static global separation can fulfill both
$W\le-M$ and $M+W\ge0$: $W=-M$, obtained by crossing $E$ with
$E_{lim}$ at the radius where $M=-W$, as shown on Fig. \ref{fig:Global-analysis-yielding}.
Note that the energy conditions does prevent initial conditions with
inward going initial flow in the neighbourhood of the global separation
if the no shell crossing condition is to be maintained. In that case
it is allowed to have initial radius outside $r_{\star}\le r_{i}$
but then the shells just outside the separating one should be ingoing
and unbound. This would result in shell crossing after some time in
a symmetric way as found in \cite{LeDMM09a} in the case of the analysis
of a $\Lambda$LTB model.

\subsection{Cracking phenomenon of Herrera and coworkers}

We find a second illustrative example of our results in the concept
of cracking put forward by Herrera and collaborators \cite{HerreraSources}
whereby a static spherical configuration is unstable to anisotropic
perturbations and \textquotedbl{}cracks\textquotedbl{}.

In order to discuss this concept within our framework we have to consider
the set of EFEs as in Sec. \ref{sub:dynamical-system-of}'s Eqs. (\ref{gRayGen})
to (\ref{eq:WeylConstraintSimple}).

From these equations we see that the shear plays a central role. From
the shear propagation equation (\ref{eq:shearRicci-1}) we realize
that if the shear were to vanish initially, any deviations from constant
curvature given by the term $\epsilon-(\Sigma-\frac{\kappa^{2}}{2}\Pi)$
would indeed make the shear become non-vanishing at any later instant.

To recover the cracking phenomena envisaged by Herrera and collaborators
we start assuming a static, isotropic, shear-free initial configuration.
Thus we put $\Theta=0$, $\Pi=0$ and $a=0$ in some region of the
initial hypersurface where we assume also that $\rho$ and $P$ only
vary slowly. Then we can assess any future deviation from that configuration
using the restriction to this hypersurface of Eqs. (\ref{gRayGen})
to (\ref{eq:WeylConstraintSimple}). The governing initial equations,
for $\Lambda=0$, $\Theta=0$, $\Pi=0$ and $a=0$, take the form
\begin{align}
\mathcal{L}_{n}\Theta= & \frac{1}{\alpha}D^{k}D_{k}\alpha-\frac{\kappa^{2}}{2}\left(\rho+3P\right),\label{eq:RayStatic}\\
\mathcal{L}_{n}a= & \epsilon-\Sigma,\label{eq:shearStatic}\\
\mathcal{L}_{n}\left(\Sigma+\frac{\kappa^{2}}{2}\Pi\right)= & 0,\label{eq:WeylStatic}\\
\Sigma= & q,\label{eq:RicciTlStatic}\\
\frac{^{3}R}{2}= & \kappa^{2}\rho\;,\label{eq:RicciTrStatic}
\end{align}
\vspace{-0.8cm}
\begin{align}
\left(P-2\Pi\right)^{\prime}= & -\left(\rho+P\right)\frac{\alpha^{\prime}}{\alpha},\label{eq:PgradStatic}\\
\left(\frac{\Theta}{3}+a\right)^{\prime}= & 0,\label{eq:ExpGradStatic}\\
\frac{\kappa^{2}}{6}\rho^{\prime}= & -\frac{\left(\left[q+a\,\left(\frac{\Theta}{3}+a\right)\right]r^{3}\right)_{\Pi=a=\Theta=0}^{\prime}}{r^{3}}\nonumber \\
= & -\frac{\left(qr^{3}\right)^{\prime}}{r^{3}}\;.\label{eq:rhoGradStatic}
\end{align}

The form of the Eq.(\ref{eq:rhoGradStatic}) uses the constraint (\ref{eq:RicciTlStatic},
actually \ref{eq:EFEricciConst}) while the Raychaudhuri Eq.(\ref{eq:RayStatic})
comes from the restriction of Eqs.(\ref{gRayGen}-\ref{eq:shearRicci-1})
to the initial configuration. Further using Eq.(\ref{eq:WeylStatic})
with the constraint (\ref{eq:EFEricciConst}) in the derivative, one
can deduce the relations between the values and proper time evolutions
of the electric Weyl scalar, the anisotropic stress and traceless
hypersurface curvature, as well as traceless Hessian scalar, shear
and expansion on the initial hypersurface
\begin{align}
-2\mathcal{L}_{n}\Sigma= & \kappa^{2}\mathcal{L}_{n}\Pi,\label{eq:Lie2shearStatic-1}\\
\mathcal{L}_{n}q= & 0\\
\Sigma= & -q,\\
a= & \Theta=\Pi=0,
\end{align}
where, in addition, the traceless Hessian scalar proper time evolution
can be obtained with the derivative of the shear Eq. (\ref{eq:shearRicci-1})
on the hypersurface
\begin{align}
\mathcal{L}_{n}^{2}a= & \mathcal{L}_{n}\left(\epsilon+\kappa^{2}\Pi\right)=\mathcal{L}_{n}\left(\epsilon-2\Sigma\right).\label{eq:Lie2shearStatic}
\end{align}
From all this, the following can be deduced: (i) the perfect fluid
source and Hessian combination in Eq.(\ref{eq:RayStatic}) drives,
in general, the expansion away from 0; (ii) combining Eqs. (\ref{eq:shearStatic})
with (\ref{eq:RicciTlStatic}), the shear, in general, is also driven
away from 0 by the difference of traceless Hessian and curvature of
the hypersurface, \textbf{$q$}, the latter mirroring the Weyl curvature,
unless they are set equal, thus\textbf{ }implying an imposed shear-free
flow; (iii) anisotropy as well is driven away from 0, in parallel
with the evolution both of the electric part of the Weyl, \textbf{$\Sigma$},
and of the difference of traceless Hessian and hypersurface curvature,
$q$, as\textbf{ }we have
\begin{align}
\mathcal{L}_{n}\Pi= & \frac{\mathcal{L}_{n}^{2}a-\mathcal{L}_{n}\epsilon}{\kappa^{2}},\label{eq:AnisoStressEvol}
\end{align}
except if the flow is restricted to being shear-free and geodesic;
(iv) the separation scalar defined in Eq. (\ref{eq:LimitDef}) is
0 on the initial hypersurface but will be driven away by
\begin{align*}
\mathcal{L}_{n}\left(\frac{\Theta}{3}+a\right)= & -q-\frac{\kappa^{2}}{2}\left(\rho+3P\right).
\end{align*}
Therefore, if a shell where $q=-\frac{\kappa^{2}}{2}\left(\rho+3P\right)$
exists in the considered region, it satisfies locally the TOV equilibrium
condition, where forces balance, and is thus surrounded by shells
experiencing nonzero forces. That shell satisfies the conditions (\ref{eq:LimitDef})
and (\ref{eq:gTOV0}) of a dividing shell. As a consequence, we realize
that $\mathrm{gTOV}$ becomes non vanishing in the neighborhood of
the dividing shell, inducing the appearance of the radial force responsible
for the cracking phenomena under the following conditions: for $\Theta_{\star}r_{\star}^{\prime}>0$,
the radial balance of the separation scalar will drive neighbouring
shells to the cracking condition of outer shells and inner shells
experiencing positive, resp. negative areal expansion, as seen in
Eq. (\ref{eq:TrjnhEFE:MomentumConstr}), at some later time from those
initial conditions. In conclusion, the cracking shell is a kind of
separating shell.

We can further extend our interpretation of cracking in this framework
by considering shear-free flows. Then, (i) expansion is still driven
away from 0 by\textbf{ $\frac{1}{\alpha}D^{k}D_{k}\alpha-\frac{\kappa^{2}}{2}\left(\rho+3P\right)$;}
(ii) anisotropy is still driven away from 0 by the traceless Hessian;
(iii) as seen in Sec. \ref{sub:Dynamical-system-restricted}, the
shearfree condition entails from Eq. (\ref{eq:constraint_2-shearfree})
that the expansion should be of uniform sign in all spacetime. The
departure from initial vanishing expansion by Eq. (\ref{eq:RayStatic})
will give its definite sign and thus the expansion is always either
all collapsing or all expanding, as found in \cite{Herrera:2010ed}.

An important point to be emphasized at this stage is that our analysis
draws on the full set on non-linear equations and is therefore more
general than that of the original works of Herrera an collaborators.
Alternatively, a perturbative gauge invariant treatment of this issue
can be done using the formalism developed by \cite{Ellis:1989jt}
and subsequently explored by others \cite{Bruni:1992dg,Challinor:1998aa,Zibin:2008vj,Sussman:2008kr,Clarkson:2009sc}.
We will address it elsewhere.\textbf{}%

We conclude this section emphasising the importance of both the shear
and the anisotropic stresses not only for the existence of a dividing
shell, but also for the cracking phenomena of Herrera and collaborators.
In the latter case, this confirms their claims in an alternative way.


\section{\label{sec:Summary-and-discussion}Summary and discussion}

In the present work we have considered spherically symmetric, inhomogeneous
universes with anisotropic stresses in order to investigate the existence
and stability of a dividing shell separating expanding and collapsing
regions. With this endeavour we have gone one step further than in
a previous work by considering a  more realistic scenario where the
matter is no longer a perfect fluid. 

This shows up to be quite important in characterizing the contrasting
dynamical behaviours of separate regions. This is relevant in relation
with the present understanding of structure formation as the outcome
of gravitational collapse of overdense patches within an overall expanding
universe, since there is an underlying expectation that the two disparate
behaviours decouple. This issue, is also related to the assessment
of the influence of global physics on local physics.

In the present work we have addressed this issue by resorting to an
ADM 3+1 splitting, utilising the so-called Generalized-Painlevé-Gullstrand
coordinates as developed in Refs.~\cite{Adler et al 05,LaskyLun06b}.
This enables us to follow a non-perturbative approach and to avoid
having to consider the matching of the two regions with the contrasting
behaviours. We have found local conditions characterising the existence
of a dividing shell which generalises our previous conditions for
perfect fluids \cite{MLeDM09}. One is a condition establishing the
precise balance between two energy quantities that are the analogues
of the total and potential energies at the dividing shell. (This amounts
to the vanishing of the kinetic energy of the shell.) The second condition
establishes that a generalized TOV equation is satisfied on that shell,
and hence that this shell is in equilibrium, but one which now involves
explicitly the anisotropic stress.\textbf{ }Moreover, the former condition
also implies that there is no matter transfer across the separating
shell, and hence we may call the region enclosed by the latter a trapped
matter region. The trapped matter is not in static equilibrium in
contrast to the situations where the TOV equation is satisfied in
the whole of the trapped region, and which are meant to describe stars\textbf{.}

We have also related these conditions to a gauge invariant definition
of the properties of the dividing shell. These require the vanishing
of a combination of the expansion scalar and of the shear, on the
shell. %
Finally, if we demand that the dividing shell is static, in order
to define an extreme case of reference, we obtain an additional equation
of state that relates the gauge-invariant Hessian trace of the model
with the quantity $\rho+3P$ involved in the strong energy condition,
on the dividing shell. Naturally in this limit case, the expansion
and shear will vanish on the shell.%

The approach followed in this paper has allowed us to translate the
Einstein field equations %
in terms of non-local quantities, such as the Misner-Sharp mass $M$
and the energy/curvature parameter $E$, as well as into equations
involving local quantities%
. The latter are convenient to describe the evolving behaviours separated
by the dividing shell. This procedure led us to relate the energy
equation to the generalized Friedmann equation, and likewise we relate
the generalized TOV equation with the generalized Raychaudhuri equation.
Moreover, we present both the equations governing the flow behaviour
of the remaining quantities such as the shear, the electric part of
the Weyl tensor, as well as the constraint equations that hold for
them and for\textbf{ }their radial gradients. We also give the constraints
and evolution applied on the expansion and shear combination. This
allowed us to discuss the dynamical behaviour in the neighborhood
of the separating shell. In particular we have obtained the condition
for a shear-free flow that generalizes previous results \cite{Mimoso:1993ym,Coley:1994yt,Herrera:2010ed}.

We have considered two illustrations of our results, namely we have
analysed the existence of a separating shell in the class of matter
and radiation solutions put forward by Sussman and Pavón. We showed
that, in this case, the existence of a dividing shell requires that
the radiation exerts a repulsive role. And we have shown that our
results allow a discussion of the emergence of the cracking phenomena
put forward by Herrera and collaborators.\textbf{ }We have described
cracking initial conditions, their dynamics, and showed, within our
gauge invariant formalism, how shear and anisotropic stress trigger
the phenomenon of cracking%
. Our approach also opens windows on the behaviours of the electric
part of the Weyl, the quantities characterising the 3-curvature%
. We also recover the properties discussed in \cite{Herrera:2010ed}
in shearfree flows.

In this paper we didn't do a thorough discussion of all dynamical
possibilities offered by the system we discovered. This opens many
future work.


\begin{acknowledgments}
The authors wish to thank José Fernando Pascual %
for helpful discussions. MLeD also wishes to thank Michele Fontanini,
Daniel Guariento and Elcio Abdalla for helpful discussions. The work
of MLeD has been supported by CSIC (JAEDoc072), CICYT (FPA2006-05807)
in Spain and FAPESP (2011/24089-5) in Brazil. %
FCM thanks CMAT, Univ. Minho, for support through FEDER Funds COMPETE
and FCT Projects Est-C/MAT/UI0013/2011, PTDC/MAT/108921/2008 and CERN/FP/116377/2010.
MleD and JPM acknowledge the CAAUL's project PEst-OE/FIS/UI2751/2011.
JPM also wishes to thank FCT for the grants %
CERN/FP/123615/2011 and CERN/FP/123618/2011. 
\end{acknowledgments}
\appendix

\section{Metric ADM scalar functions}

For clarity, we present the scalar gauge invariants involved in the
ADM formulation in terms of the GPG metric functions, starting with
the perfect fluid terms
\begin{align}
\Theta= & 2\frac{\mathcal{L}_{n}r}{r}-\frac{\beta^{\prime}}{\alpha}-\frac{1}{2}\frac{\mathcal{L}_{n}E}{1+E}\nonumber \\
= & -\frac{1}{\alpha r^{2}}\left(r^{2}\beta\right)^{\prime}-\frac{1}{2}\frac{\mathcal{L}_{n}E}{1+E}+\frac{2\dot{r}}{\alpha r},\\
a= & \frac{1}{3}\left[\frac{\mathcal{L}_{n}r}{r}+\frac{\beta^{\prime}}{\alpha}+\frac{1}{2}\frac{\mathcal{L}_{n}E}{1+E}\right],\nonumber \\
= & \frac{1}{3}\left[\frac{r}{\alpha}\left(\frac{\beta}{r}\right)^{\prime}+\frac{1}{2}\frac{\mathcal{L}_{n}E}{1+E}+\frac{\dot{r}}{\alpha r}\right],
\end{align}
which, combined with Eq. (\ref{eq:LieE}), yield
\begin{align}
\Theta= & \frac{\left(\mathcal{L}_{n}r\right)^{\prime}}{r^{\prime}}+2\frac{\mathcal{L}_{n}r}{r},\label{eq:ThetaInRj}
\end{align}
\begin{align}
a= & -\frac{1}{3}\left[\frac{\left(\mathcal{L}_{n}r\right)^{\prime}}{r^{\prime}}-\frac{\mathcal{L}_{n}r}{r}\right].\label{eq:ShearInRj}
\end{align}
 
\begin{align}
^{\left(3\right)}R & =-\frac{2}{r^{2}}\left[\left(\left(1+E\right)\left(r^{2}\right)^{\prime}\right)^{\prime}-r^{\prime}\left(\left(1+E\right)r\right)^{\prime}-1\right]\nonumber \\
 & =-2\left\{ \left(1+E\right)\left(\frac{r^{\prime}}{r}\right)^{2}-\frac{1}{r^{2}}+2\frac{\sqrt{1+E}}{r}\left(\sqrt{1+E}r^{\prime}\right)^{\prime}\right\} \label{eq:3Rdef}\\
 & =-\frac{2}{r^{2}}\left\{ \left(Err^{\prime}\right)^{\prime}+\left(1+E\right)rr^{\prime\prime}+\left[\left(rr^{\prime}\right)^{\prime}-1\right]\right\} ,\\
q & =\frac{1}{6}\left\{ r\left(\frac{Er^{\prime}}{r^{2}}\right)^{\prime}+E\frac{r^{\prime\prime}}{r}+\frac{2}{r^{2}}\left[1+r^{2}\left(\frac{r^{\prime}}{r}\right)^{\prime}\right]\right\} \nonumber \\
 & =\frac{1}{6}\left[r\left(\frac{Er^{\prime}}{r^{2}}\right)^{\prime}+\left(2+E\right)\frac{r^{\prime\prime}}{r}+\frac{2}{r^{2}}\left(1-r^{\prime2}\right)\right],
\end{align}
\begin{align}
\frac{1}{\alpha}\, D^{\mu}D_{\mu}\alpha & =\frac{\sqrt{1+E}}{\alpha r^{2}}\left(r^{2}\sqrt{1+E}\alpha^{\prime}\right)^{\prime},\\
\epsilon & =-\frac{r\sqrt{1+E}}{3\alpha}\left(\frac{\sqrt{1+E}}{r}\alpha^{\prime}\right)^{\prime},
\end{align}
and from the shear evolution comparing that from the EFE, Eq.(\ref{eq:evol2}),
and that obtained from the Ricci identities, we get
\begin{align*}
\Sigma= & \frac{3\kappa^{2}}{2}\Pi-q-a\left(\frac{\Theta}{3}+a\right).
\end{align*}





\end{comment}
\end{thebibliography}

\end{document}